# Moving Magnetic Flux and Electromagnetic Induction

Allan Walstad
Physics Department
University of Pittsburgh at Johnstown
awalstad@pitt.edu

Abstract: Magnetic flux can move, and moving magnetic flux is associated with induced electric fields. Where there are multiple sources of flux, the induced electric field is the vector resultant of the induced fields attributable to the fluxes acting individually; it is not necessarily associated with the resultant magnetic field. The key insight is that flux can move through regions in which the magnetic field itself vanishes, much as electric charge can move (as electric current) through a region in which the net charge density is zero. For this reason, moving flux can account for the induced emf around a long solenoid in which the current is changing. It also accounts for unipolar induction, and this account is testable.

## 1. Introduction

Maxwell, in summarizing Faraday's research into electromagnetic induction, captures the understanding that will be elaborated in what follows [1]:

> The conception which Faraday had of the continuity of the lines of force precludes the possibility of their suddenly starting into existence in a place where there were none before. [We understand this in terms of the Maxwell equation $\nabla \cdot \vec{B} = 0$.] If, therefore, the number of lines which pass through a conducting circuit is made to vary, it can only be by the circuit moving across the lines of force, or else *by the lines of force moving across the circuit*. In either case a current is generated in the circuit. [Emphasis added.]

In modern parlance, Faraday's "lines of force" (which he also referred to as "magnetic curves") are field lines, a pictorial representation of both field and flux. We take each line to represent an equal quantity of flux. The field strength at any location is represented by the number of lines per area through a small surface normal to the lines. With this interpretation, Faraday's view (as described above by Maxwell) envisions electromagnetic induction as arising from motion relative to magnetic flux, whether wires are moving through a static field or the flux itself is moving.

We illustrate this viewpoint by considering a horseshoe magnet with broad north and south poles facing each other across an air gap. We move the magnet from one side to the other side of a lab table. The magnetic flux, which remained concentrated between the poles, moved across the table with the magnet. Suppose that stationary metal rings were located at each side of the table, one intercepting the magnet's flux at the start, the other intercepting the magnet's flux at the end. Then, magnetic flux traveled out of the first ring and into the second, accounting by Faraday's Law for a transient induced emf and current in each ring.

Emf in a stationary ring requires the existence of an electric field to exert forces on conduction charges in the ring. Therefore, in the example above, moving magnetic flux is associated with an induced electric field. Could it be that *all* induced electric fields are associated with moving magnetic flux? The results of this paper suggest that the answer is "yes," as advocated long ago by Poynting [2]:

> Whenever electromotive force is produced by change in the magnetic field, or by motion of matter through the field, the E.M.F. per unit length or the electric intensity is equal to the number of tubes of magnetic induction cutting or cut by the unit length per second, the E.M.F. tending to produce induction in the direction in which a right-handed screw would move if turned round from the direction of motion relatively to the tubes toward the direction of magnetic induction.

Poynting also sought to treat *magnetic* fields as induced by the motion of *electric* flux (tubes of induction), an idea that will not be taken up here.

Nevertheless, by now the idea that flux, or field lines, or the field itself can move through space appears to be the object of skepticism in physics. Feynman [3] dismissed the concept in his influential *Lectures*. Page and Adams [4] warned that "[t]he concept of moving tubes of induction is one which should be avoided as it often leads to erroneous conclusions." Purcell [5] allowed only that the *magnetic field pattern* (Purcell's italics) moves with a moving magnet, possibly leading to a change in flux through a wire loop and thereby an induced emf. A paper by Roche [6] allowed for moving lines in cases like our example above, where a single magnet undergoes translational motion, but not where electromagnetic induction is the result of a changing current. Good [7] conceded that "it is sometimes intuitively attractive to imagine field lines sweeping through space or cutting through conductors" but asserted nevertheless that "fields don't move", that the idea of a moving field is "physically meaningless", that the magnetic field lines of a current-carrying wire are "*not* being carried along with the wire" (Good's italics). Nevertheless, Taylor and Leus [8] have made effective use of the concept in the case of translational motion of the source. Most of the literature has centered on unipolar induction and the question of whether the field lines of a permanent magnet spinning about its polar axis rotate with the magnet. On this topic, Baumgartel and Maher [9], in a paper describing their own recent experiments, provide over 100 references to previous theoretical and experimental work, and McDonald has posted online a seemingly exhaustive reference list [10]. An apparent obstacle to any *general* association of moving flux with induction has been the existence of induced electric fields in regions of zero or near-zero magnetic field, such as outside a long solenoid with changing current. There, electromagnetism texts appeal to the vector potential. In so doing, Lorrain et al. [11] comment that "'flux-cutting' is not an entirely satisfactory interpretation of the Faraday induction law." In the same context Shadowitz [12] asserts that "$\vec{A}$ has direct physical significance while $\vec{B}$ is, more or less, a mathematical aid." Several private conversations and internet searches have reinforced my impression of a prevailing skepticism toward the concept of moving flux. The purpose of this paper is to allay that skepticism.

In section 2 we look more closely at the induced emf in a wire ring due to its motion relative to the poles of a horseshoe magnet. The force on a charged particle in the ring depends only on the relative velocity (for velocities much less than that of light). If the particle is at rest, this force is attributed to an induced electric field, for which a simple expression obtains. In section 3 we consider a circular current loop with increasing current, and we find that the induced electric field at some distance from the loop is quantitatively associated with outward-moving magnetic flux. Section 4 presents a key insight, that magnetic flux can move through a region of *zero* magnetic field, with an associated electric field in the region. More generally, when there are multiple sources of flux, the induced electric field is the vector resultant of the induced fields attributable to the fluxes acting individually; it is not necessarily associated with any presumed motion of the resultant magnetic field. For this reason, when the current in a long solenoid is changing, the induced emf in a surrounding ring is attributable to moving magnetic flux from the solenoidal turns, even though the net magnetic field itself is near zero in the vicinity of the ring. Section 5 notes that the emf around a real-world (finite-length) solenoid can be associated by Faraday's Law with changing magnetic flux through a closed path that does not enclose the interior solenoidal flux. Section 6 treats unipolar induction, demonstrating that the emf generated by a permanent magnet spinning about its axis of symmetry is associated quantitatively with the moving fluxes of magnetic dipoles comprising the magnet. Section 7 argues for the value of diverse viewpoints from which to treat physical phenomena. Section 8 re-states the principal contributions of the paper and suggests that an invariable association of induced electric fields with moving magnetic flux would justify viewing the association as causal: induced electric fields are caused by moving magnetic flux.

## 2. Motion relative to magnetic flux

In figure 1a, a wire loop of height $h$ traveling at speed $v$ enters the uniform magnetic field $B_0$ between the rectangular poles of a broad-faced horseshoe magnet. Within the field region there is a magnetic force-per-charge $f = vB_0$ on conduction charges in the wire, leading to an emf $\mathcal{E}_0 = vB_0h$ around the loop.

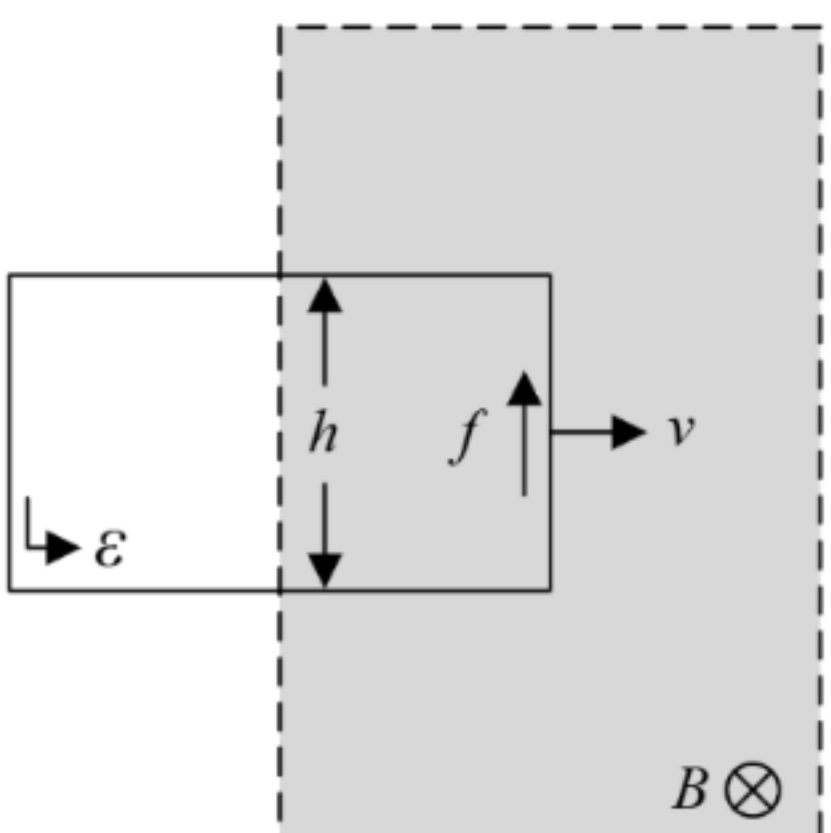

Figure 1a. A rectangular wire loop traveling at speed $v$ enters the uniform magnetic field between the poles of a broad-faced horseshoe magnet. The magnetic force-per-charge on conduction charges in the wire results in an emf around the loop.

In figure 1b, the loop is stationary but the magnet is moving in the opposite direction at speed $v$. It seems there can be no magnetic force on the stationary conduction charges, but by Lorentz transformation of the field tensor the pure magnetic field in the rest frame of the magnet leads to an electric field in the rest frame of the loop (and of the lab). This electric field has magnitude $E = \gamma v B_0$, where $\gamma = (1 - v^2/c^2)^{-1/2}$ is the Lorentz factor, and it points in the same direction as f previously. The result is an emf $\varepsilon_L = \gamma v B_0 h$ in the rest frame of the loop. Since the magnetic field $B_L$ in the rest frame of the loop is, by transformation of the field tensor,[1] equal to $\gamma B_0$, we have $\varepsilon_L = v B_L h$.

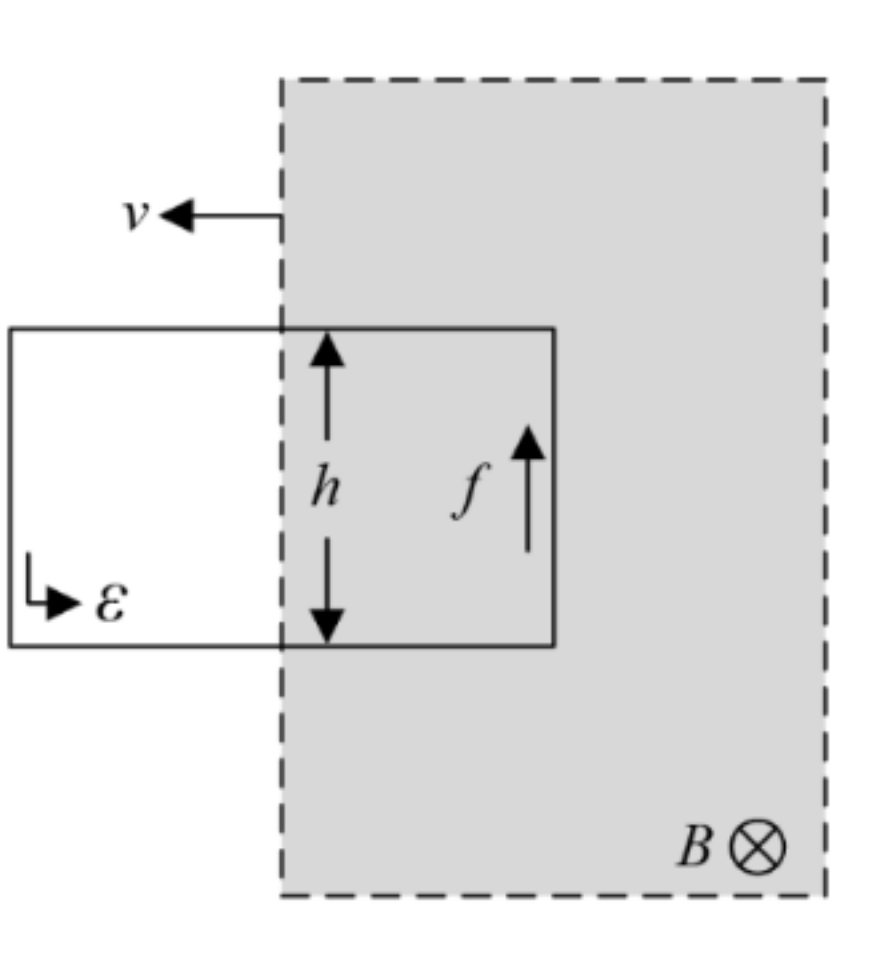

Figure 1b. Similar to figure1a except that the loop is stationary and the magnet is moving.

Since $vB_0h$ or $vB_Lh$ is the rate, respectively, at which flux enters the loop in figure 1a or figure 1b, we see that in both frames—that of the magnet and that of the loop--the induced emf is equal in magnitude to the rate of change of magnetic flux Φ through the loop,

$$|\varepsilon| = \left|\frac{d\Phi}{dt}\right|, \qquad (1)$$

with the sense of the emf given by Lenz's law. Equation (1) is Faraday's law.

Excepting a factor of $\gamma$, the magnetic field is the same in both frames, as is the force on a conduction charge. Henceforth we will take those equalities as given, leaving our results throughout this paper reliably accurate to first order in $v/c$, which will be suitable for treating typical induction experiments in the laboratory. Then the relevant expression for the magnetic force on a conduction charge is not just q$vB$, where $v$ is the velocity of the charge, but q$v_{rel}B$ where $v_{rel}$ is the velocity of the charge *relative to the flux*. In both of the two basic cases, or in the case where both loop and magnet are moving, $v_{rel}Bh$ is the rate at which magnetic flux is entering the loop.

If the relative motion is along the direction of the field, then no force is exerted on conduction charges. Hence, wire must cut across flux (or, equivalently, flux must cut across wire) for emf to be generated. By analyzing the relative velocity vector into components along and perpendicular to $\vec{B}$, we see that the force on the charged particle is given by

$$\vec{F} = q\vec{v}_{rel} \times \vec{B}. \quad (2)$$

If $\vec{v}_q$ is the velocity of the particle and $\vec{v}_f$ the velocity of the flux, then the relative velocity is $\vec{v}_{rel} = \vec{v}_q - \vec{v}_f$. If the particle is at rest, or in any case where our interest lies specifically in the force due to the motion of magnetic flux relative to the observer's rest frame, equation (2) permits us to define an induced electric field as

$$\vec{E} = \vec{B} \times \vec{v}_f. \quad (3)$$

Could we have said "motion of the *field*"? That is, in figure 1b, it may be tempting to say that the magnetic field moves with the magnet. Nevertheless, the magnetic field has long been understood as flux density: the directed flux per area through an infinitesimal area oriented so as to maximize the flux through it. The movement of flux into or out of an area can change the flux density, i.e., change the field in that region. It is sufficient to say only that flux moves. An analogy may be helpful. As charge flows into or out of a region, the charge density in that region can change, but we would probably not say that the charge density moves. Charge is what moves. Since, as we shall see, flux can move through a region of zero field (just as charge can move [as current] through an electrically neutral wire), in this paper we speak only of moving flux or motion relative to flux.

## 3. A circular current loop with increasing current

Figure 2 shows two circular loops of wire, the inner of which has cross-sectional area $a$ and current $I$ that is increasing at the rate $dI/dt \equiv \dot{I}$. At any moment, the directed flux, as illustrated in familiar diagrams by field lines, passes upward through this loop and wraps around, passing in the opposite direction through the plane of the loop outside. As the current increases, so does the total flux. Since flux wraps around the current producing it, new flux must crowd inward within the loop while simultaneously expanding outward on the outside[2] in order that the magnetic field reaches its final configuration once the current stops increasing. The induced electric field generates an emf and current in the outer wire loop.

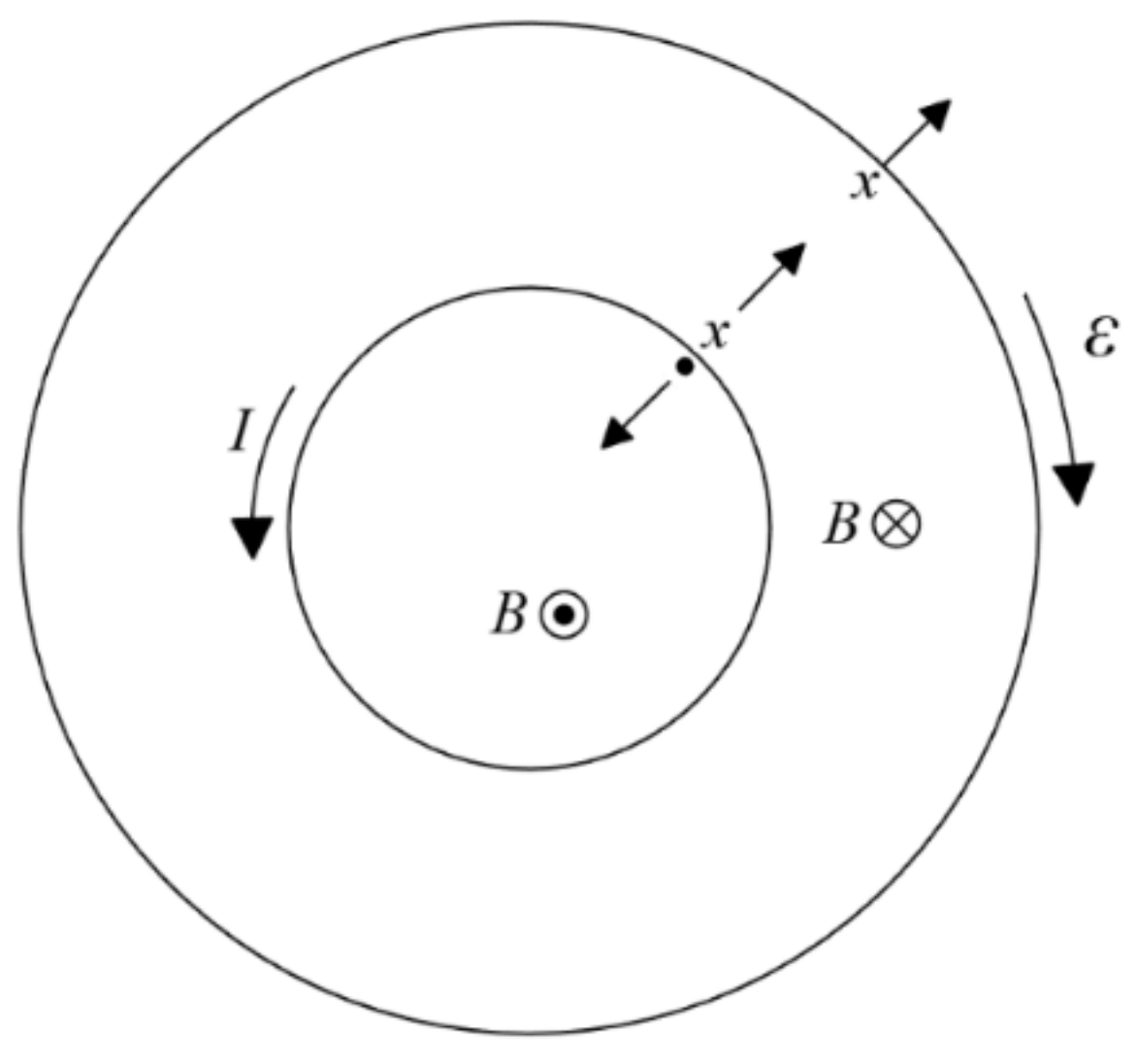

Figure 2. An increasing current in a small wire loop generates outward-moving magnetic flux that is intercepted by a larger, concentric loop. The induced electric field generates an emf.

We are interested in the resulting electric field not just in the plane of the current loop but at distance $r$ in any direction $\theta$ from its axis. We take $r$ to be much greater than the loop radius. We will find the correct expression using the vector potential, then demonstrate that the same expression is expected from moving flux. The vector potential is given (neglecting the small propagation time[3]) by

$$\vec{A} = \frac{\mu_0}{4\pi}\frac{Ia}{r^2}\sin\theta\ \hat{\phi}, \tag{4}$$

where $\hat{\phi}$ denotes the azimuthal direction. The electric field is

$$\vec{E} = -\frac{\partial \vec{A}}{\partial t} = -\frac{\mu_0}{4\pi}\frac{\dot{I}a}{r^2}\sin\theta\,\hat{\phi} \tag{5}$$

and the magnetic field is

$$\vec{B} = \vec{\nabla}\times\vec{A} = \frac{\mu_0}{4\pi}\frac{Ia}{r^3}\left(\sin\theta\,\hat{\theta} + 2\cos\theta\ \hat{r}\right). \tag{6}$$

We see as follows that $\vec{E}$ is attributable to magnetic flux moving outward from the loop at a velocity $\vec{v}_f = v_f\hat{r}$. Consider a cone about the axis of the loop, taken to be the z axis, opening from the origin by the angle $\theta$. (If $\theta$ were equal to $90^0$ then the "cone" would be the plane of the current loop.) As the current in the loop increases, so does the total flux outward through the curved surface of the cone, and in particular so does the total flux $\Phi_{>r}$ beyond some radial distance $r$. We have

$$\dot{\Phi}_{>r} = \frac{d}{dt}\int_r^{\infty} 2\pi r \sin\theta \, B_\theta \, dr = \frac{\mu_0 \dot{I} a}{2r}\sin^2\theta \ . \qquad (7)$$

At distance r, outward-moving flux crosses a circle of circumference $2\pi r \sin\theta$. The rate of increase of $\Phi_{>r}$ is related to the outward speed by

$$\dot{\Phi}_{>r} = 2\pi r B_\theta \, v_f \sin\theta. \qquad (8)$$

From equations (6)-(8) we find

$$v_f = \frac{\dot{I}}{I} r \ . \qquad (9)$$

(If the current has been increasing uniformly from zero, we can write this as $v_f = r/\tau$ where $\tau = I/\dot{I}$ is the time elapsed since the start. Then $v_f$ would reach the speed of light $c$ at a distance $r = c\tau$ which is the maximum distance which the outward-moving front of magnetic flux can have reached in time $\tau$.) Finally, by equations (6) and (9) we can write

$$\vec{E} = \vec{B} \times \vec{v}_f = -\frac{\mu_0}{4\pi}\frac{\dot{I}a}{r^2}\sin\theta \, \hat{\phi}, \qquad (10)$$

which agrees with equation (5).

The above result applies just as well if the current is decreasing; in that case, flux travels inward toward the loop.

## 4. Flux moves through a region of zero field; implications

Figure 3 depicts a side view of two identical, broad-faced magnets resting close together with their north poles opposed. The field of each has the same magnitude B, but the resulting field in the region between the poles is zero. (Both sets of [oppositely directed] field lines are indicated in figure 3.) Suppose that the bottom magnet is moving to the right at speed $v \ll c$. A (positively) charged particle sits in the middle of the region between the poles. Is there a force on the particle?

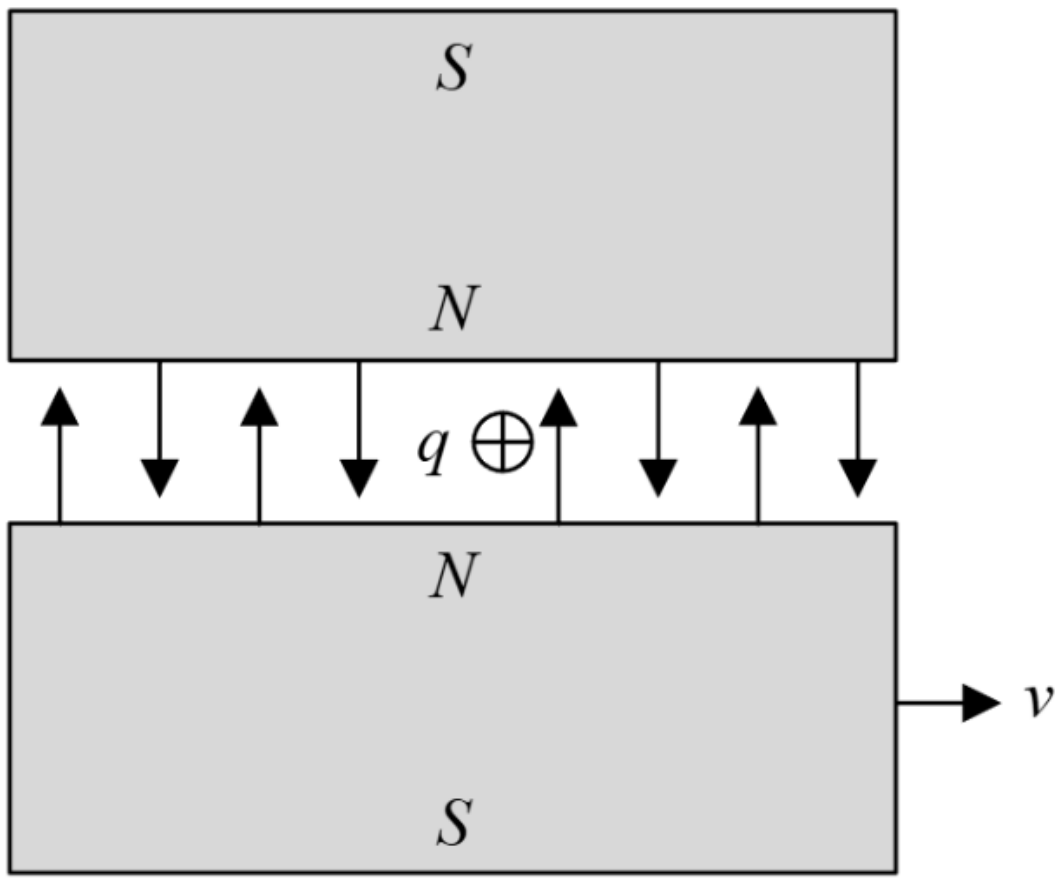

Figure 3. Two identical broad-faced magnets face each other with a small space in between. The bottom magnet is moving to the right.

There is, because the moving flux of the bottom magnet is associated with an induced electric field given by $\vec{E} = \vec{B} \times \vec{v}_f$ which points into the page. There is no electric field associated with the top magnet, since its flux remains at rest. (The conventional explanation is that the pure magnetic field of the bottom magnet in its own rest frame, when the field tensor is transformed to the lab frame, results in an electric field pointing into the page. This electric field is not affected by the presence of the stationary top magnet.)

The same conclusion must apply whenever the flux from multiple sources has different velocities: the induced electric field does not derive from any motion of the flux associated with the net magnetic field. Rather, each individual component of flux, traveling at its velocity, contributes independently to $\vec{E}$. Put succinctly, the electric field due to moving flux is not obtained from

$$\vec{E} = (\textstyle\sum_i \vec{B}_i) \times \vec{v}_f \quad \text{(wrong)}$$

but from

$$\vec{E} = \textstyle\sum_i \left(\vec{B} \times \vec{v}_f\right)_i \; . \tag{11}$$

Poynting incorporated the same insight, allowing for the independent motion of "induction tubes" oriented in different directions [13].

Equation (11) is the basis for understanding how moving magnetic flux can account for the electric field and emf around a long solenoid, even though the magnetic field in the vicinity of the solenoid is very nearly zero. Given that the induced electric field of the solenoid is the resultant of the electric fields induced by each of its turns, and having seen in section 3 that the electric field induced by the changing current of any single turn can be understood in terms of moving flux from that turn, the matter is settled: the emf around the solenoid is attributable to moving flux. The same conclusion applies straightforwardly to the case of a toroidal solenoid. Back-emf is understandable similarly, in that the moving flux from each turn acts to induce emf in every other turn.

## 5. Emf around a solenoid is associated with changing magnetic flux *outside* the solenoid

In figure 4, a long solenoid is viewed from one end, its internal magnetic field pointing out of the page. As the solenoidal current increases, the magnetic flux $\Phi$ inside increases at a rate $d\Phi/\mathrm{dt}$, and a clockwise emf of that magnitude is detected in a circular loop of radius $r$. We are assuming that $r$ is very small compared to the length $L$ of the solenoid. For simplicity and definiteness, we also assume that the circle lies in the midplane of the solenoid.

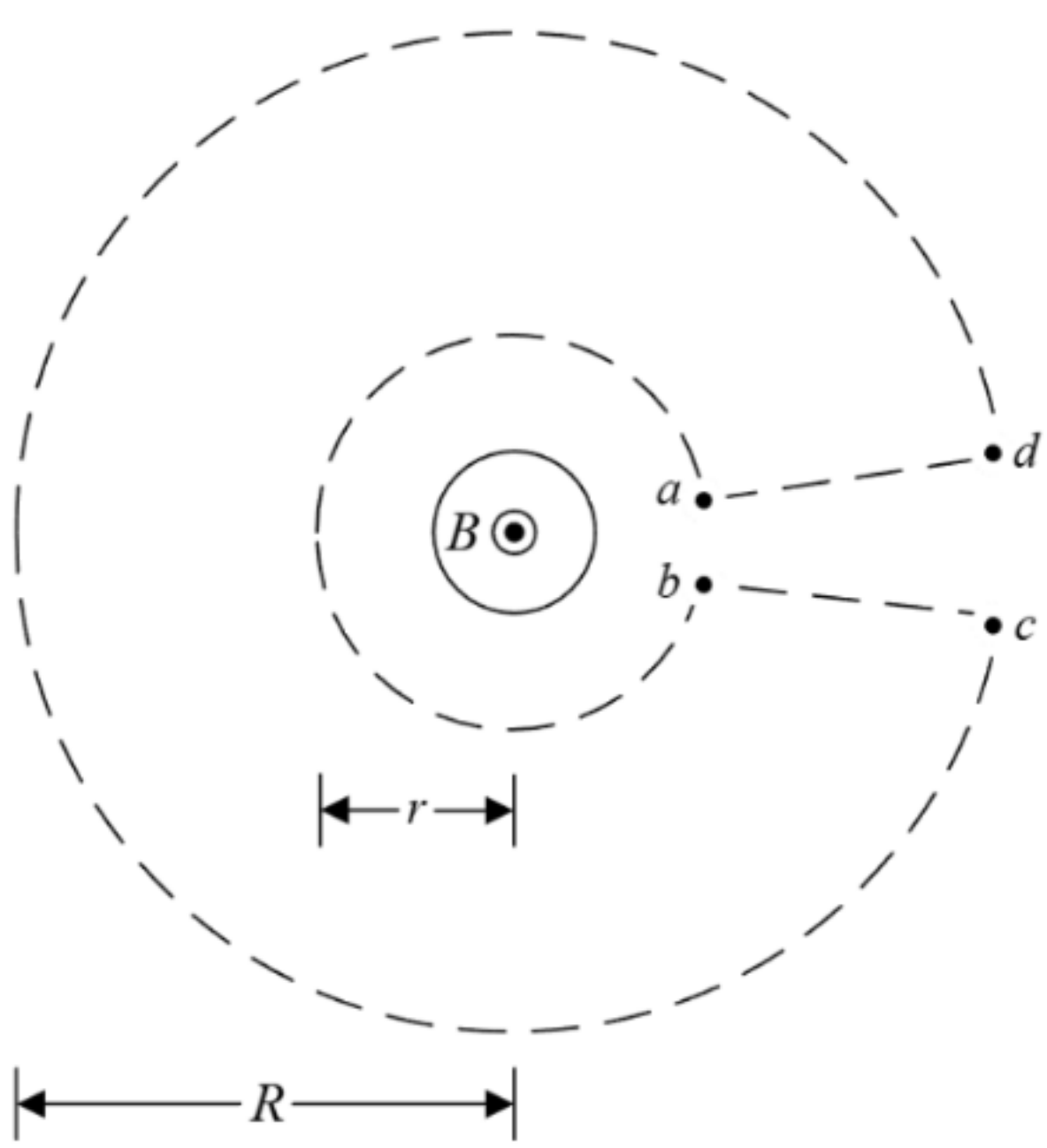

Figure 4. The closed loop $abcd$ does not enclose the flux within the solenoid, which is viewed from the north end.

Let us allow for an infinitesimal break (exaggerated in figure 4) in the circle between points $a$ and $b$, leading by radial segments to a larger circle of radius $R$. The result is a loop, extending counter-clockwise from $a$ to $b$, radially out to $c$, clockwise from $c$ to $d$, and radially back to $a$. This loop does NOT contain the solenoidal flux.

If $R$, like $r$, is much smaller than $L$, a clockwise emf is induced in the larger circle, equal to that in the smaller circle, and, since any contributions from the radial segments would cancel, we see that the emf around the loop $abcd$ is zero. Hence, by Faraday's law, there is no changing magnetic flux through the loop $abcd$.

Nevertheless, there are no infinitely long solenoids. For $R \gg L$, the emf induced around the large circle tends to zero. Then there *is* an emf around the loop $abcd$, which by Faraday's law requires a changing magnetic flux through that loop. But the emf around the loop $abcd$ is just the emf around the small circle. Thus, emf induced around a long solenoid is attributable via Faraday's law to changing magnetic flux in the region *outside* the solenoid. We now have a conceptually appealing and quantitatively accurate model of how that happens: magnetic flux propagates outward from the solenoid and cuts the loop, accounting locally via $\vec{E} = \vec{B} \times \vec{v}_f$ for the electric field and resulting emf.

## 6. The spinning magnet and unipolar induction

Unipolar induction is illustrated in figure 5. When the cylindrical permanent magnet rotates about its axis, an emf and current are induced in the circuit that is completed via sliding contacts with the side and axis of the magnet. (We make the usual assumption that the current path within the magnet extends along the axis from the pole, then radially from the axis to the side contact.) We will see that the emf cannot be explained on the basis that the flux associated with the net magnetic field *rotates* with the magnet. Rather, it is associated with the *translational* motion of the fluxes of many individual magnetic dipoles in the magnet, each acting independently.

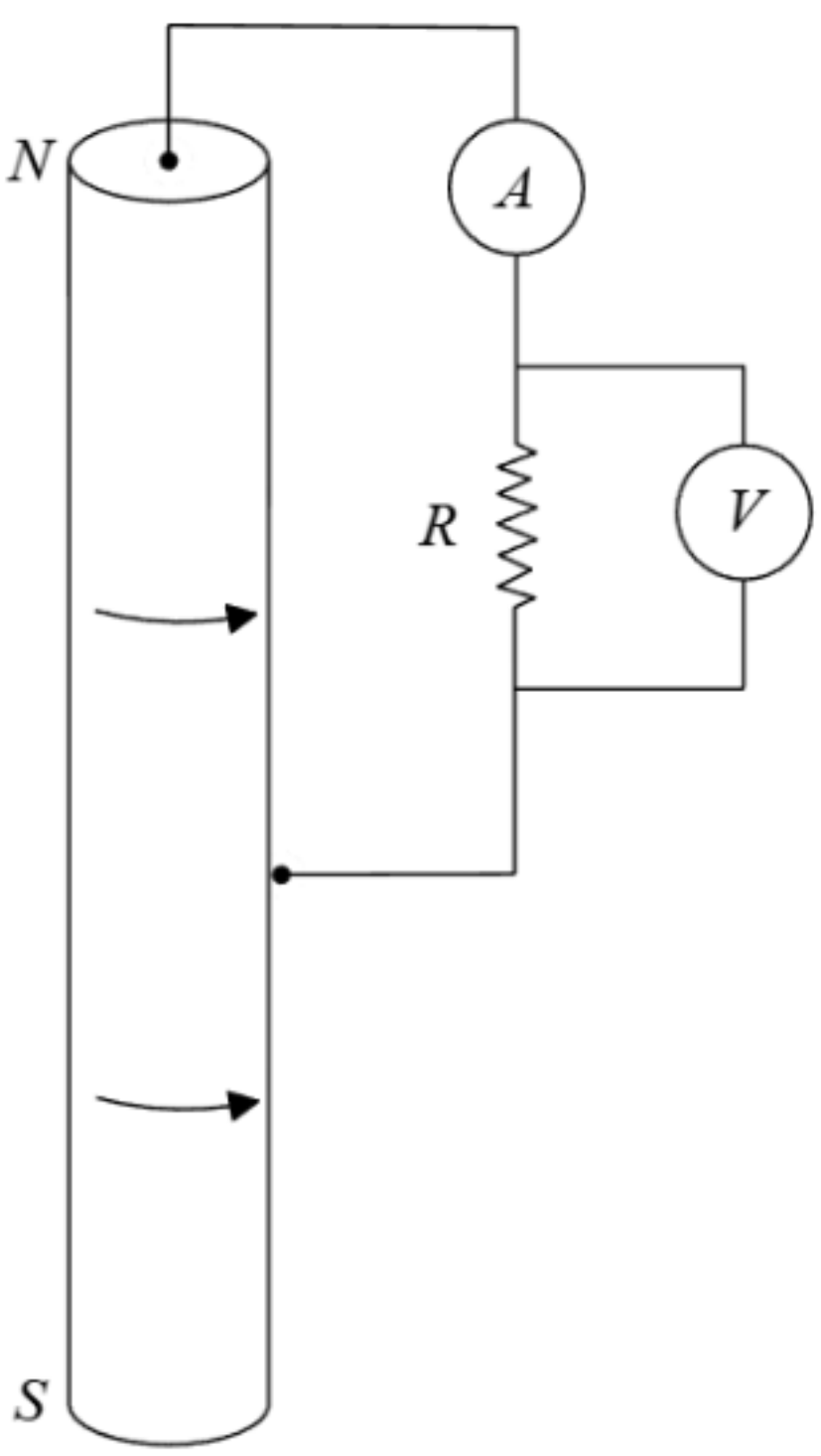

Figure 5. A unipolar generator.

We address the question of rotation by considering a circular current loop that rotates clockwise about its axis with its current flowing also in the clockwise direction. At the surface of the wire, the magnetic field and flux wrap around it such that, if the flux were rotating with the loop, the induced electric field given by $\vec{E} = \vec{B} \times \vec{v}_f$ would point outward from the surface. By Gauss's law, the wire would have a net positive charge. As charge is conserved, however, a neutral object cannot acquire a net charge simply by being placed into rotation. We must conclude that the magnetic flux of a circular current loop rotating about its axis does not rotate with the loop.

A permanent magnet comprises many small magnetic dipoles, which in classical electromagnetism we may reasonably model as circular current loops. In the situation depicted in figure 5, although the flux from each dipole does not rotate[4], it is undergoing translational motion (excepting dipoles located on the magnet's axis). We expect, therefore, an induced electric field that is the resultant of the induced electric fields ascribable to the motions of the many component dipoles. Over a century ago, Swann [14] affirmed the validity of this method in a paper intended to determine "the extent to which the 'moving line theory' is the equivalent of the Maxwell-Lorentz theory." (Before proceeding, is emphasized that although the following analysis is presented in terms of moving flux, the same results follow from transformation of the pure magnetic field of each moving current loop, in its own rest frame, to the rest frame of the laboratory, yielding an electric field and emf.)

To perform the sum, we first divide the magnet into many infinitesimally thin, parallel filamentary magnets. Since all the sources in one such filament share a common velocity, we may apply $\vec{E} = \vec{B} \times \vec{v}_f$ to the resultant magnetic field *of that filament*, which is identical to the magnetic field of a long thin solenoid. It comprises a uniform internal field and an external field attributable to a north magnetic monopole at one end and an equal south monopole at the other. Therefore, the magnetic field of our spinning magnet (being the sum of the fields of the many filamentary solenoids) consists of a uniform internal field, added to the fields of a north monopole disc at one end and a south monopole disc at the other. Analytically, we organize the monopoles into concentric rings.

The induced electric field of a monopole moving with velocity $\vec{v}$ is easy to find. Denoting its flux by $\Phi$, the magnetic field at location $\vec{r}$ is

$$\vec{B} = \frac{\Phi}{4\pi r^3}\, \vec{r}. \tag{12}$$

Then we have

$$\vec{E} = \vec{B} \times \vec{v}_f = \frac{\Phi}{4\pi r^3}\, \vec{r} \times \vec{v}\ . \tag{13}$$

If the velocity lies along the $+x$ direction, and if $\vec{r}$ points at an angle $\alpha$ to that direction, then $\vec{r} \times \vec{v}$ is equal to r $v \sin\alpha$ in the direction opposite to the direction that fingers of the right hand would wrap around the $x$ axis with thumb in the $+x$ direction. That is, excepting a factor of $\Phi/\mu_0 q$, the induced electric field associated with a moving north monopole is identical to the *magnetic* field that would be associated with an oppositely-moving positive *electric* point charge q.

It follows that a narrow, uniform circular *ring* of monopoles, possessing total flux $\Phi$, is associated with an electric field identical (excepting the factor of $\Phi/\mu_0 Q$) to the magnetic field of a charged ring, of charge Q and radius r, rotating about its axis in the opposite direction with tangential velocity $v$. The current in this case is

$$I = \frac{Qv}{2\pi r}. \tag{14}$$

Then the electric field of the rotating monopole ring is related to the magnetic field of a circular current loop by

$$\vec{E} = \frac{\Phi}{\mu_0 Q}\vec{B} = \frac{\Phi v}{2\pi r \mu_0 I}\vec{B}. \tag{15}$$

Of special interest to us is the resulting emf $\varepsilon$ around a closed path that passes once through the area of the ring:

$$\varepsilon = \oint \vec{E} \cdot d\vec{\ell} = \frac{\Phi v}{2\pi r \mu_0 I}\oint \vec{B} \cdot d\vec{\ell} = \frac{\Phi v}{2\pi r \mu_0 I} \cdot \mu_0 I = \frac{\Phi v}{2\pi r}. \tag{16}$$

Since the tangential velocity $v$ is related to the angular velocity $\omega$ of the monopole ring by $v = \omega r$, we obtain for the induced emf of the spinning dipole ring

$$\varepsilon = \frac{\Phi\omega}{2\pi}. \tag{17}$$

The end of a cylindrical magnet spinning about its axis can be modeled as a uniform disk comprised of monopole rings. We wish to find the resultant emf around the circuit of figure (5). Now letting $\Phi$ be the *total* flux from the magnet's pole and $R$ be the radius of the disk, the flux associated with a ring of radius $r$ and thickness $dr$ is

$$d\Phi = \frac{2\Phi}{R^2} r\, dr. \tag{18}$$

The contribution to the emf from the ring is by equation (17)

$$d\varepsilon = \frac{\omega}{2\pi} d\Phi = \frac{\omega\Phi}{\pi R^2}\, r\, dr \tag{19}$$

which integrates to

$$\varepsilon = \frac{\omega\Phi}{2\pi}. \tag{20}$$

It should come as no surprise that equation (20) is identical to equation (17), with the total flux replacing the ring flux, since $\varepsilon$ for each ring depends only on its flux and angular velocity, not its radius. The circuit of figure 5 is completed in the rotating material of the magnet, presumed conducting. At each point, the material has no velocity relative to the filament passing through that point, the internal field of which therefore does not contribute to induced emf in the circuit. Nor do the internal fields of other filaments act at that point. Nevertheless, there is one additional source. Since magnetic field does not rotate, the conducting material itself is rotating through the magnetic field of the monopole disc. If the cylindrical magnet is much longer than wide, this field becomes negligible within the body of the magnet far from its poles. Then, the emf in figure 5 is given accurately by equation (20). Close to one of the poles, the Lorentz force on conduction charges in the magnet results in an emf that opposes the emf of equation (20). Hence, in general, the emf is strongest when the circuit is connected near the midpoint of the magnet (as in figure 5) and decreases as the connection point moves closer to the pole.

We have, then, a theory of unipolar induction in which moving flux plays a central role. Let us call this "theory A."

The same results can derived without the concept of moving flux by assuming that the emf arises entirely within the magnet from the motion of the magnetic medium in its own (static) magnetic field $B_{med}$ as a consequence of the Lorentz force. Let us call this "theory B." The radial force-per-charge on conduction charges in the medium at a distance r from the axis is

$$f = vB_{med} = \omega r B_{med}. \quad (21)$$

Far from the poles, the field within the magnetic medium is given simply by

$$B_{med} = \Phi/\pi R^2. \quad (22)$$

Substituting equation (22) into equation (21) and integrating from the axis to the outer edge of the magnet gives an expression for emf identical to equation (20) in the situation of figure 5. As $B_{med}$ decreases near the poles, theory B continues to give the same emf as theory A. Nevertheless, theories A and B disagree fundamentally with regard to the origin of the emf.

It appears that we can distinguish between them via an experiment similar to one described [15] by Guala-Valverde, who was a proponent of rotating magnetic fields. Figure 6 shows a long rotating cylindrical magnet from which a thin wedge has been removed. A conducting probe extends from the axis out to a conducting ring, with which it makes a sliding contact. A wire from the ring completes a circuit in which voltage is measured. The voltage reading is compared to the reading obtained with an otherwise identical magnet having no wedge or radial probe, where the circuit is connected as in figure 5.

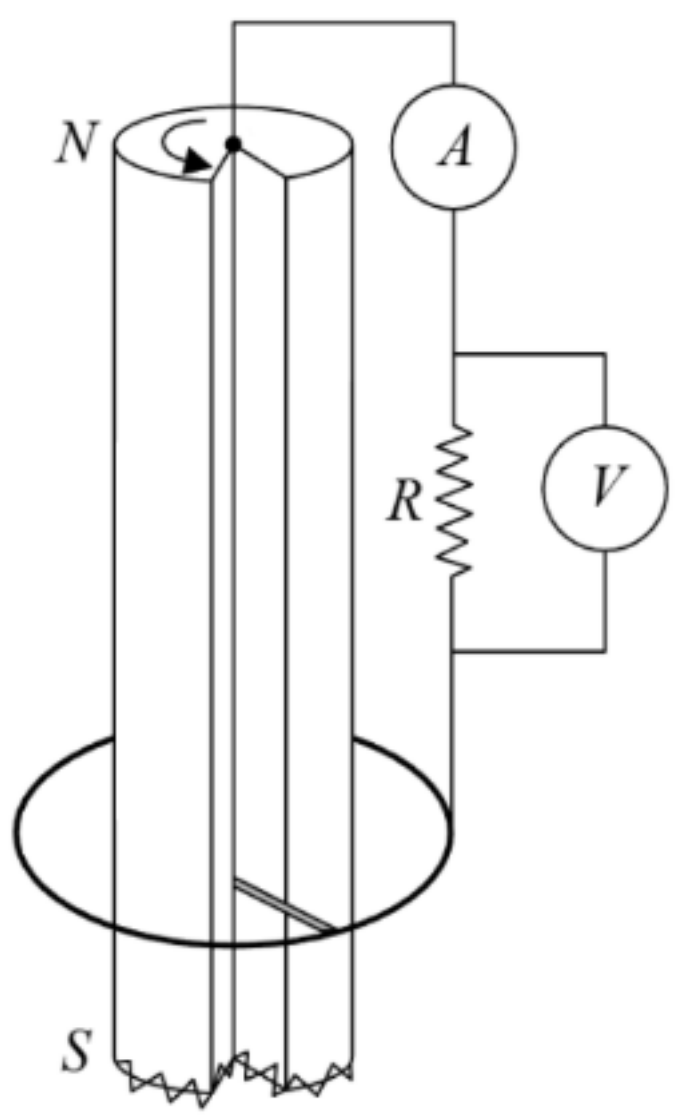

Figure 6. Unipolar generator missing a wedge. A probe extends from the axis to a stationary ring, with which it makes a sliding contact.

According to theory A, the readings will be nearly identical, because the emf was due entirely to the spinning monopole disc, and the thin wedge will make only a small deduction. According to theory B, the readings will be entirely different, because the emf is claimed to be due to the motion of a conducting material through a magnetic field, but the field in the wedge region is negligible. Guala-Valverde's result was incompatible with theory B (and apparently dispositive in ruling it out), but the experiment described here is not identical to Valverde's, in which the probe could either rest on top of a thin disk magnet or lie within an empty wedge. The experiment described in this paper would be needed to confirm theory A.

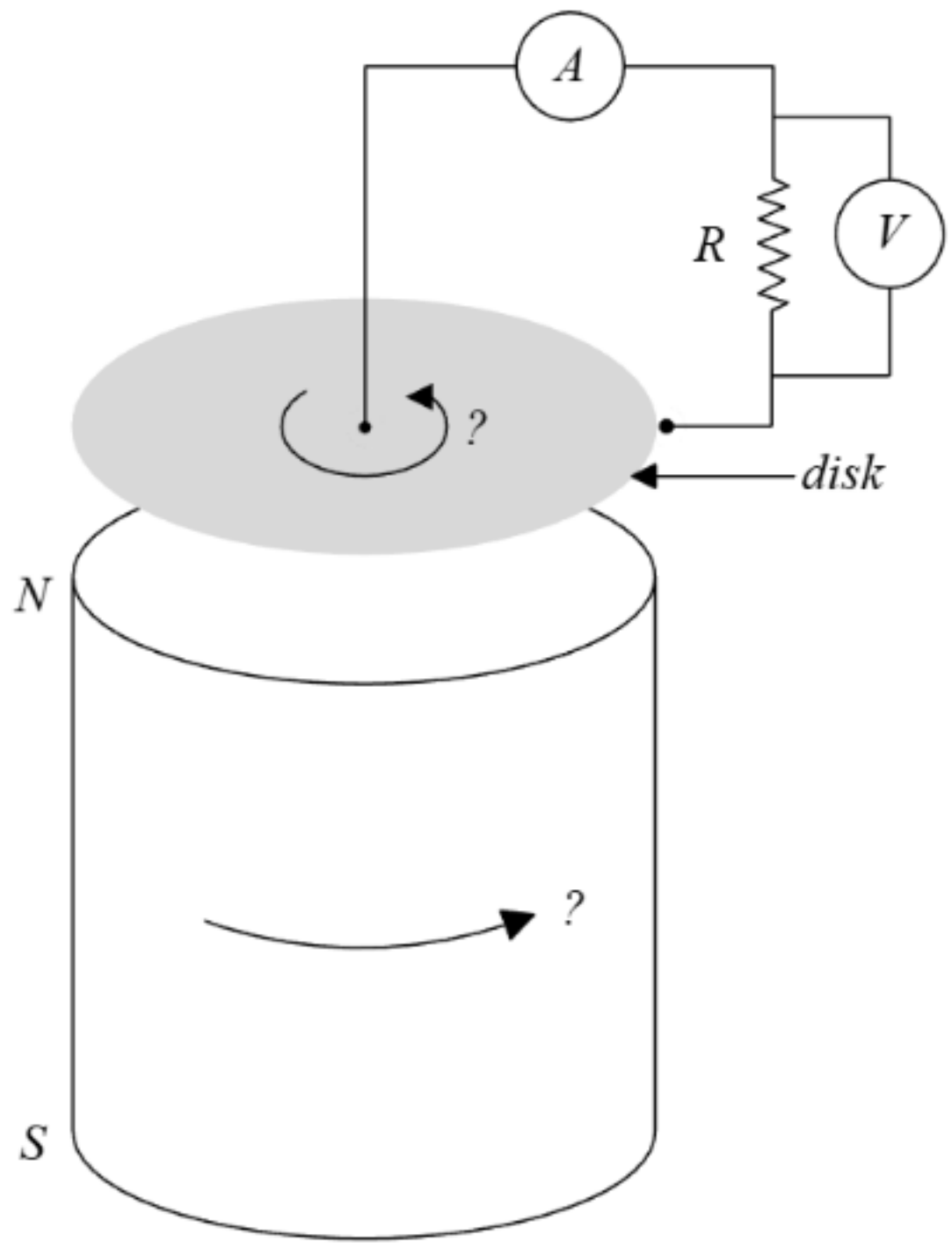

Figure 7. The Faraday disk.

A variant of the unipolar generator is the Faraday disk. A non-magnetic conducting disk is mounted beyond the end of a cylindrical magnet and an emf is measured as depicted in figure 7. Several combinations of rotation may be implemented. The magnet alone may be set into rotation; the disk alone may rotate, with the circuit completed via sliding contacts; the magnet and disk may rotate together; the entire system including the meters and wires may rotate together; etc. Because the magnetic field does not rotate, and because the circuit in which emf is measured lies entirely outside the magnet, the rotation of the magnet itself has no bearing on the results. That is, the measured emf in every case must be the result of the disk and/or the wires rotating in a static magnetic field.

## 7. Editorial

To a student contemplating the scenario described in the introduction to this paper, it must seem obvious that the flux between the poles of the magnet moved with it across the lab table, thereby leaving or entering a metal ring and inducing emf via Faraday's law. We ought not discourage the student's impression. Moving magnetic flux is a valid conceptual model of electromagnetic induction.

Throughout physics, we have access to multiple perspectives from which to address specific problems. For example, classical mechanics is rooted in Newton's laws, but we often apply them through principles such as the conservation of momentum or angular momentum or energy, or through the principle of least action and Lagrange's and Hamilton's equations. Different viewpoints best serve different purposes. With regard to electromagnetic induction, we often, but not always, calculate induced emf via Faraday's law. In our example of the moving magnet, not only do we see how moving flux enters and exits a metal ring, but we can understand the origin of the emf in the induced electric fields associated with moving flux. Where, in the frame of our laboratory, it is the metal ring that is moving, or in the case of the spinning Faraday disk, we see the Lorentz force law acting on conduction charges moving through a magnetic field. For the solenoid with changing current, while moving flux remains valid, the vector potential seems a more apt approach to understanding how the changing flux through the solenoid results in an electric field outside. With regard to the unipolar generator, it may be that moving flux offers the crucial insight. Physics students will benefit from appreciating the value of multiple models rooted in the same physics, and from adding to their repertoire the model that has been validated in this paper.

## 8. Conclusion

The association of induced electric fields with moving magnetic flux has faced an apparent counterexample in the long solenoid, where a changing current induces an electric field in a surrounding region of near-zero magnetic field. Nevertheless, as demonstrated in Sec. 4 above, magnetic flux can propagate through a region of zero magnetic field, much as charge can flow through a neutral medium, the moving flux from multiple sources acting independently in electromagnetic induction. With this insight, the emf induced around a long solenoid is accounted for by moving magnetic flux. Unipolar induction is also accounted for by moving magnetic flux, and it appears possible experimentally to distinguish this account from a plausible alternative.

If, as now seems likely, induced electric fields are always associated quantitatively with moving magnetic flux, it becomes reasonable to promote that association to a causal model of electromagnetic induction: induced electric fields are induced by--caused by--moving magnetic flux.

## Acknowledgment

The author thanks Hannah Y. Mannion for expert preparation of the diagrams.

## Footnotes

[1] The factor of $\gamma$ can be seen as arising from the motion of the magnetic flux: Lorentz contraction compresses the flux into a smaller area, and field is flux per area.

[2] The newly created flux could not instantly appear in its final distribution at all distances from the loop without violating causality. It must therefore travel outward from the source. Poynting [2] likewise recognized that flux ("induction tubes") must move in response to a changing source current.

[3] We are assuming as a matter of approximation that the effects of a changing current are transmitted instantly in low-frequency laboratory experiments on induction. There is no conflict with footnote 2, which refers to a matter of principle.

[4] That is, if we represent the magnetic field of the dipole by an array of field lines, the orientation of the array remains fixed in space as the dipole orbits the rotational axis of the magnet.

**References**

[1] Maxwell J C 1954 *A Treatise on Electricity and Magnetism, Unabridged Third Edition, Vol. II* (Dover) p 189

[2] Poynting J H 1885 On the connexion between electric current and the electric and magnetic inductions in the surrounding field *Philosophical Transactions of the Royal Society of London* **176** 280

[3] Feynman R P, Leighton R B, and Sands M 1964 *The Feynman Lectures on Physics, Vol. II* (Addison-Wesley) pp 13:10-11

[4] Page L and Adams N I 1949 *Electrodynamics, 2nd Ed.* (Van Nostrand) p 329

[5] Purcell E M 2011 *Electricity and Magnetism, 2nd Ed.* (Cambridge) p 270

[6] Roche J 1987 Explaining electromagnetic induction: a critical re-examination *Phys. Educ.* **22** (2) 91-99

[7] Good R H 1999 *Classical Electromagnetism* (Saunders) p 105

[8] Taylor S and Leus V 2012 The magneto-kinematic effect for the case of rectilinear motion *Eur. J. Phys.* **33** (4) 837-852

[9] Baumgartel C and Maher S 2022 Resolving the paradox of unipolar induction: new experimental evidence on the influence of the test circuit *Scientific Reports* **12** 16791

[10] McDonald K T: Is Faraday's disk dynamo a flux-rule exception? kirkmcd.princeton.edu/examples/faradaydisk.pdf

[11] Lorrain P, Corson D R, and Lorrain F 2000 *Fundamentals of Electromagnetic Phenomena* (W. H. Freeman) p 333

[12] Shadowitz A 1988 *The Electromagnetic Field* (Dover) p 389

[13] Poynting [2] p 295.

[14] Swann W F G 1920 Unipolar induction *Phys. Rev.* **15** (5) 365-398. The relevant statement is on p 385. I am indebted to Kirk McDonald for alerting me to Swann's paper.

[15] Guala-Valverde J 2002 On the electrodynamics of spinning magnets *Spacetime & Substance* **3** 140-144